\documentclass[preprint,12pt]{aastex}
\usepackage{natbib}
\begin{document}

\title{Constraints on the equation of state of dark energy \\ and the
Hubble constant from stellar ages and the CMB\altaffilmark{1}}

\author{Raul Jimenez\altaffilmark{2}, Licia Verde\altaffilmark{3,4},
Tommaso Treu\altaffilmark{5}, and Daniel Stern\altaffilmark{6}}

\altaffiltext{1}{Partly based on observations collected at ESO (La
Silla and Paranal) under programmes 62.O-0592, 63.O-0468, and
64.O-0281, 65.O-0446 and 66.A-0362, and at the W.~M. Keck Observatory,
which is operated jointly by the California Institute of Technology and
the University of California.}

\altaffiltext{2}{Department of Physics \& Astronomy, University of
Pennsylvania, Philadelphia, PA 19104. Email:  raulj@physics.upenn.edu}

\altaffiltext{3}{Department of Astrophysical Sciences, Princeton
University, Princeton, NJ 08544-1001. Email:
lverde@astro.princeton.edu}

\altaffiltext{4}{Chandra fellow, Spitzer fellow}

\altaffiltext{5}{California Institute of Technology, Astronomy,
mailcode 105-24, Pasadena CA 91125. Email: tt@astro.caltech.edu}

\altaffiltext{6}{Jet Propulsion Laboratory, California Institute of
Technology, Mail Stop 169-327, Pasadena, CA 91109. Email:
stern@zwolfkinder.jpl.nasa.gov}

\begin{abstract} 

We place constraints on the redshift-averaged, effective value of
the equation of state of dark energy, $w$, using {\em only} the
absolute ages of Galactic stars and the observed position of the first
peak in the angular power spectrum of the CMB.  We find $w<-0.8$ at the
68\% confidence level. If we further consider that $w \geq -1$, this
finding suggests that within our uncertainties, dark energy is
indistinguishable from a classical vacuum energy term.

We detect a correlation between the ages of the oldest galaxies and
their redshift. This opens up the possibility of measuring $w(z)$
by computing the relative ages of the oldest galaxies in the universe
as a function of redshift, $dz/dt$. We show that this is a realistic
possibility by computing $dz/dt$ at $z \sim 0$ from SDSS galaxies and
obtain an independent estimate for the Hubble constant, $H_0 = 69 \pm
12$ km s$^{-1}$ Mpc$^{-1}$.  The small number of galaxies considered at
$z>0.2$ does not yield, currently, a precise determination of $w(z)$,
but shows that the age--redshift relation is consistent with a Standard
$\Lambda$CDM universe with $w=-1$.

\end{abstract}


\section{Introduction}

There is now solid observational evidence that the universe is
accelerating. The absolute ages of high-redshift galaxies \citep{D+96,
Spinrad+97} and the value of the Hubble constant $H_0$
\citep{Freedman+01}, the Hubble diagram of Type~Ia supernovae
\citep{Riess+98, Perlmutter+99}, the angular power spectrum of the
cosmic microwave background \citep[CMB;][]{Benoit+02, Ruhl+02,
Goldstein+02} in combination with galaxy surveys
\citep{Efstathiou+2dF02, Verde+2dF02} and lensing \citep{WTJZ02}, all
indicate that the universe is flat and dominated at present by some
form of dark energy with negative pressure. Strong confirmation for
this comes from the newly released results from the {\it Wilkinson
Microwave Anisotropy Probe} ({\it WMAP}\,) first year data
\citep{SpergelWMAP03}.

The equation of state of the dark energy, $p=w \rho$, expresses the
ratio between the pressure, $p$, and the mass density, $\rho$, of the
dark energy in terms of the parameter $w$ (in units of $c=1$).  The
value of $w$ could either be constant, as in the case of a cosmological
constant ($w=-1$), or be time-dependent, as in the case of a rolling
scalar field or ``quintessence'' \citep{PeeblesRatra88, CDS98}. Any
such behavior would have far-reaching implications for particle
physics.

In this paper we use the ages of the oldest stellar populations between
$z=0$ and $z=1.5$ to derive new and independent constraints on two key
cosmological parameters, $w$ and $H_0$.  Both parameters are relatively
poorly constrained by CMB data alone.  For example, in order to get
accurate constraints on $w$, \citet{SpergelWMAP03} combined the
{\it WMAP} data with six external data sets:  two additional CMB data
sets \citep[CBI and ACBAR;][]{Pearson+02, Kuo+02}, large scale
structure measurements from the 2dF galaxy redshift survey
\citep[2dFGRS;][]{Percival+01, Verde+2dF02}, measurements of the
Lyman~$\alpha$ forest power spectrum \citep{Croft+02, Gnedin+02}, the
luminosity distance--redshift relation from Type~Ia supernovae
measurements \citep{Garnavich+98, Riess+01}, and $H_0$ from the {\it
Hubble Space Telescope} ({\it HST}\,) Key Project \citep{Freedman+01}.
Therefore it is particularly important to obtain accurate and
independent measurements of cosmological parameters relying on
different physics.

In Section~2 we focus on $w$, summarizing the basis of our method and
deriving tight constraints on the equation of state using the ages of
globular clusters in combination with the location of the first
acoustic peak in the CMB power spectrum. Then, in Section~3, we
determine stellar ages for a sample of early-type galaxies between
$z=0$ and $z=1.5$ and use the age--redshift relationship defined by the
old envelope to determine $H_0$.  The prospects of using this method to
derive a measurement of $w(z)$ are also discussed.  Section 4
summarizes and discusses the results.

Throughout the paper we adopt the following notation. The Hubble
constant is $H_0=100\,h\,$ km s$^{-1}$ Mpc$^{-1}$ and $c$ is the speed
of light. The density of baryons, matter, radiation, photons,
neutrinos, and dark energy, at present, in critical units, are
indicated as $\Omega_b$, $\Omega_m$, $\Omega_{\rm rad}$,
$\Omega_{\gamma}$, $\Omega_{\nu}$, and $\Omega_{\Lambda}$ respectively
($\Omega_{\rm rad}\equiv\Omega_{\gamma}+\Omega_{\nu}$). The physical
quantities, independent of $h$, are indicated with $\omega$; e.g.,
$\omega_m=\Omega_mh^2$. When needed, a standard, flat, reference
$\Lambda$CDM model with $\Omega_m=0.27$, $h=0.71$, $\Omega_b h^2=0.023$,
and $w=-1$ is assumed \citep{SpergelWMAP03}.

\section{The value of $w$ from stellar ages}

In this section we use the ages of the oldest globular clusters in
combination with the location of the first acoustic peak of the CMB
power spectrum \citep{PageWMAP03} to constrain the value of $w$. The
method is briefly described in Section 2.1 and the results are given
in Section 2.2.

\subsection{Method}

If the universe is assumed to be flat, the position of the first
acoustic peak ($\ell_1$ in the standard spherical harmonics notation)
depends primarily on the age of the universe and on the effective
value\footnote{I.e., the average value over redshift. We use
$w(z)$ to indicate when $w$ is allowed to vary over time.} of $w$
\citep{CDS98,HFZT01,KCS01}.  As noted by these authors, for a fixed $w$
value, a change in the physical density parameter $\omega_m=\Omega_mh^2$
that keeps the characteristic angular scale of the first acoustic peak
fixed will also leave the age approximately unchanged. Thus an independent
estimate of the absolute age of the universe at $z=0$ combined with a
measurement of $\ell_1$ yields an estimate of $w$ largely independent
of other cosmological parameters (e.g., $\Omega_m$ and $h$).

This can be better understood by considering that for a constant $w(z)$,
the age of a flat universe is given by
\begin{equation}
t_0= H_0^{-1} \int_{0}^{\infty} (1+z)^{-1}\left[\Omega_{\rm
rad}(1+z)^4+\Omega_m(1+z)^3+\Omega_{\Lambda} (1+z)^{3(1+w)}\right]^{-1/2}
dz\;.
\label{eq:ageflatwconst}
\end{equation}
The position of the first acoustic peak is fixed by the quantity
\begin{equation}
\theta_A=r_s(a_{dec})/d_A(a_{dec})\;,
\label{eq:thetaa}
\end{equation}
where $a_{dec}$ is the scale factor at decoupling, $r_s(a_{dec})$
is the sound horizon at decoupling, and $d_A(a_{dec})$ is the angular
diameter distance at decoupling.  For a flat universe,
\begin{equation}
r_s(a_{dec})=\frac{c}{H_0\sqrt{3}}\int_0^{a_{dec}}\left[\left(1 +
\frac{3\Omega_b} {4\Omega_{\gamma}}\right)(\Omega_{\Lambda}x^{1-3w} +
\Omega_mx + \Omega_{\rm rad})\right]^{-1/2}dx
\label{eq:rs}
\end{equation}
and
\begin{equation}
d_A(a_{dec})=\frac{c}{H_0}\int_{a_{dec}}^1\left[\Omega_{\Lambda}x^{1-3w}
+ \Omega_mx + \Omega_{\rm rad}\right]^{-1/2}dx\;.
\label{eq:dA}
\end{equation} 
See \citet{VerdeWMAP03} for more details.  We use equations
\ref{eq:ageflatwconst} to \ref{eq:dA} and fix $\theta_A$ to be
consistent with the {\it WMAP} determination \citep[$\ell_1\simeq
220$;][]{PageWMAP03}.  Figure~1 shows the allowed region in the age-$w$
plane obtained for $\Omega_b h^2=0.02$, $0.05<\Omega_m<0.4$, and
$0.5<h<0.9$. It is clear from the plot that an independent and accurate
age determination of the universe will provide a measurement of $w$. Thus
by using only the {\it WMAP} observation of $\ell_1$ and an independent
estimate of the age of the universe, we can place constraints on $w$,
largely independent of other cosmological parameters.

\subsection{Results from the ages of globular clusters}

The ages of the oldest globular clusters (GCs)
provide a lower limit to the total age of the universe.  Numerous
star-forming galaxies have now been observed up to redshift $z=6.6$
\citep[e.g.,][]{Kodaira+03}.  Since the oldest GCs contain the oldest
stellar populations in galaxies, it is reasonable to assume that GCs too
have formed at redshift $z>6$.  Furthermore, since small halos are easily
destroyed after reionization \citep[e.g.,][]{Bullocketal01}, GCs likely
formed prior to the reionization epoch.  Under this assumption, the {\it
WMAP} recent estimate of the reionization redshift $z_{\rm reion}\sim
17$ \citep{KogutWMAP03} implies that the ages of the oldest globular
cluster do not significantly underestimate the age of the universe.

Globular cluster ages can reliably be estimated in several different
ways: from the luminosity of the main-sequence turn-off (if the
distance is known), from detached binaries, from the termination of
the white-dwarf sequence, and from the luminosity function method
which provides a joint constraint on the age and distance of a GC.
\citet{R+96} determine the distance to NGC~6752 by means of accurate
photometry of GC white dwarfs and obtained an age of $14.5 \pm 1.5$
Gyr.  Using the same method they find 47 Tucanae to have an age of $13
\pm 2.5$ Gyr \citep{Zoccali+01}.  \citet{Gratton+97}, using the new
{\it Hipparcos} distance scale to determine main-sequence turn-offs,
estimate an age of $11.8^{+2.1}_{-2.5}$ Gyr (95\% confidence level)
for the oldest Galactic GCs.  \cite{JimenezPadoan97} apply the
luminosity function method to M55 and obtain an age of $12.5 \pm 1.0$
Gyr (95\% confidence level).  \citet{Hansen+02}, using very deep {\it
HST} images, find the white dwarf cooling sequence for M4 and estimate
its age to be $12.7 \pm 0.7$ Gyr at the $95$\% confidence level.
Using the theoretical age--mass relation and applying it to the
detached eclipsing binary OGLEGC-17, \cite{ChaboyerKrauss02} determine
Omega Centauri to have an age of $11.2 \pm 1.1$ Gyr (95\% confidence
level).  These age estimates are all remarkably consistent.
\citet{KraussChaboyer03} perform the most careful analysis to date of
the effects of systematics in GC age determinations.  They estimate
the age of the oldest Galactic GCs using the main-sequence turn-off
luminosity and evaluate the errors with Monte-Carlo techniques, paying
careful attention to uncertainties in the distance, to systematics,
and to model uncertainties.  They find a best-fit age for the oldest
Galactic GCs of $12.5^{+3.4}_{-2.2}$~Gyr (95\% confidence limits).  We
adopt their probability distribution for the oldest GC age, which we
find can be accurately described by
\begin{equation}
P(t)=\frac{A}{\sigma (t-T)} \exp \left[- \frac{\ln \left[ (t-T)/m \right
]^2}{2 \sigma^2} \right ]
\label{eq:kraussagecorr}
\end{equation}
where $t$ denotes the age of the oldest GCs, $A=1.466$, $\sigma=0.25$,
$T=6.5$ and $m=5.9$. Since reionization took place at $z\sim 17$
\citep{KogutWMAP03} and we assume that oldest GCs must have formed
before then, for all reasonable cosmologies we only need add $0.2-0.3$
Gyr to the GC ages to obtain an estimate of the age of the
universe. We conservatively add 0.3 Gyr\footnote{The uncertainty in
this correction is negligible compared to the uncertainty in the
determination of GC ages.}.

In order to constrain $w$, we assume a flat universe and Monte Carlo
simulate the distribution of $\ell_1$ subject to only weak constraints
on the other cosmological parameters.  For different values of $w$, we
generate $10^5$ models, randomly sampling the cosmological parameters
$\Omega_m$, $\Omega_bh^2$ and $h$ with uniform priors, $0.5< h < 0.9$,
$0.05 < \Omega_m < 0.45$, and $0.01 < \Omega_bh^2 < 0.03$.  We then
impose an age of the universe constraint by randomly sampling these
models so that the age of the universe has a probability distribution
whose shape is given by equation \ref{eq:kraussagecorr}, but offset by
$0.3$ Gyr.  We use the publicly available code {\sf CMBFAST}
\citep{SeljakZaldarriaga96} to compute the resulting distribution of
$\ell_1$. This is shown in Figure~2 where the two solid lines are the
68\% and 90\% {\it joint} confidence levels.

As expected, the age alone does not constrain $w$; there is a degeneracy
between $w$ and $\ell_1$.  If we now use the {\em observed} position
of the first acoustic peak as recently measured, in a model independent
way, from {\it WMAP} \citep{PageWMAP03}, we can constrain $w$ with high
accuracy. The filled contours in Figure~2 are marginalized values for $w$
at the 68\% and 90\% confidence levels. Thus we find $w<-0.8$ ($w < -0.67$)
at the 68\% (90\%) confidence level.  If we drop the assumption that the
oldest GC must have formed before reionization, but still impose that
they must have formed by $z=6$, this constraint is slightly weakened. In
this case the age distribution of equation \ref{eq:kraussagecorr}
needs to be offset by $\la 1$ Gyr and we obtain $w<-0.5$ at the 90\%
confidence level.  We have also repeated the calculation using Gaussian
priors for the cosmological parameters, $h=0.7 \pm 0.2$, $\Omega_m=0.25
\pm 0.2$, $\Omega_b h^2=0.02 \pm 0.01$, and find similar, but slightly
more stringent constraints on $w$.

We stress here that this determination depends solely on the GC
determination of the age of the universe and on the {\em observed}
position of the first acoustic peak in the CMB power spectrum. Our
constraint is slightly less stringent than that obtained by
\citet{SpergelWMAP03} from a joint likelihood analysis of {\it WMAP}
with six external data sets ({\it WMAP} + CBI + ACBAR + 2dFGRS +
Lyman$\alpha$ forest power spectrum + Type~Ia supernovae + $H_0$
constraint from the {\it HST} key project), but is tighter than the
CMB-only ({\it WMAP} + CBI + ACBAR) determination and comparable to the
{\it WMAP} + ACBAR + CBI + {\it HST} constraint.

\section{$H_0$ and $w(z)$ from the ages of unresolved stellar populations
at $z=0-1.5$}

\subsection{Method}

We now turn our attention to the age of the oldest objects in the
universe at $z \ge 0$ as a function of redshift. As recently suggested
by \citet{JL02}, these can be used to determine $H(z)= - \frac{1}{1+z}
\frac{dz}{dt}$ and therefore to measure $w(z)$ and $H_0$.  The Hubble
parameter is given by $H^2(z) = H^2_0 [\rho_{\rm tot}(z) / \rho_{\rm tot}(0)]$,
where $\rho_{\rm tot}$ is the total density of the universe.  For a flat
universe composed of matter and dark energy, $H(z)$ is related to the
equation of state by
\begin{equation}
H_0^{-1}{dz\over dt} = - (1+z) {H(z)\over H_0}= -(1+z)^{5/2}
\left [\Omega_m +\Omega_{\Lambda} \exp \left \{3 \int_{0}^{z}
\frac{w(z^\prime)}{(1+z^\prime)} dz^\prime  \right \} \right ]^{1/2} ,
\label{eq0}
\end{equation}
where we have used the conservation equation for dark energy density,
$\dot{\rho_\Lambda} = -3 H(z) [1 + w(z)] \rho_\Lambda$.  Thus a
measurement of $dz/dt$ is also a measurement of $w(z)$; for further
details, see \citet{JL02}.

To obtain $dz/dt$, \citet{JL02} proposed to use the old envelope of
the age--redshift relation of E/S0 galaxies.  We assume that E/S0 at
different redshifts are drawn from the same parent population with the
bulk of their stellar populations formed at relatively high redshift
\citep[e.g.,][]{Bower+92, Stanford+98}.  Then, at relatively low redshift,
they are evolving passively and may be used as ``cosmic chronometers''.

In the rest of this section we will apply this method to a sample of
spectra (described in \S 3.2) to identify the age--redshift relation
defined by the old envelope (\S 3.3) and derive an independent
measurement of $H_0$ (\S 3.4).  We show that at the level of accuracy
of our measurement, the use of the age--redshift relation allows us to
apply this method even if a fraction of the stars in E/S0 galaxies is
assembled at $z < 1$ \citep[e.g.,][]{TSCMB02}.  We also briefly discuss
the prospects of using this method to derive $w(z)$.

\subsection{Sample and measurement}

In order to apply this method we need high-quality spectra of a sample of
old stellar populations covering the largest possible range in redshift.
To this aim we combine the following data sets: (i) the luminous red galaxy
(LRG) sample from the Sloan Digital Sky Survey (SDSS) early data release
\citep{Eisenstein+01}; (ii) the sample of field early-type galaxies from
\citet[][hereafter the Treu et al.\ sample]{TSCMB99,TSMCB01,TSCMB02};
(iii) a sample of red galaxies in the galaxy cluster MS1054$-$0321 at
$z=0.83$; and (iv) the two radio galaxies 53W091 and 53W069 \citep[][Dey
et al., in preparation]{D+96,Spinrad+97,NDJH03}.

The LRG sample is the most heterogenous, since it includes red
galaxies in general, regardless of morphology.  The Treu et al.\
sample contains high quality spectra with average $S/N=16$ per
\AA. The spectra in the Treu et al.\ sample have higher $S/N$ than the
typical LRG spectrum, but the wavelength coverage is better for the
LRG sample (see example in Figure~3). To select potentially passively
evolving galaxies, we include only galaxies with $S/N > 9$ per \AA\,
and for which passively evolving synthetic stellar population models
\citep{JPMH98} provide a good fit to the continuum. This selection
criterion discards dusty or star-forming objects. The spectra of seven
red galaxies in the cluster MS1054$-$0321 were obtained by one of us
(DS) on UT 11 March 2002 using the dual-beam Low Resolution Imaging
Spectrometer \citep[LRIS;][]{Oke+95} at Keck Observatory.  On the blue
side, we used the 300 lines mm$^{-1}$ grism ($\lambda_{\rm blaze} =
5000$ \AA; spectral resolution $\Delta \lambda_{\rm FWHM} \approx
13.5$ \AA) and on the red side, we used the the 400 lines mm$^{-1}$
grating ($\lambda_{\rm blaze} = 8500$ \AA; spectral resolution $\Delta
\lambda_{\rm FWHM} \approx 8$ \AA).  These observations, which were
obtained at the parallactic angle, totaled one hour of integration.
In order to ensure no gaps in the spectra, half the data were obtained
with a D560 dichroic and the remainder were obtained with a D680
dichroic.  These spectra have a $S/N$ of 8.  We also add the red radio
galaxies 53W091 and 53W069 for which accurate ages have already been
determined \citep[][Dey et al., in
preparation]{D+96,Spinrad+97,NDJH03}. Examples of the spectra used are
given in Figure~3.

The age of the dominant stellar population in the galaxies is obtained
by fitting single stellar population models \citep{JPMH98,JDMPP03} to
the observed spectrum.  These models have been extensively tested
\citep{KFJ02,JFMG03} and have two free parameters: age and
metallicity.  By construction, star formation is assumed to occur in a
single burst of duration much shorter ($< 10$\%) than the current age
of the galaxy \citep{JFDTPN99}. Therefore, the derived ages are
single-burst equivalent ages. The best fit model is found by standard
$\chi^2$ minimization. Then, stellar ages and their errors are obtained
by marginalizing the resulting likelihood surface over metallicity.
Some examples for the parameter confidence regions, at the 1 and
2$\sigma$ level, are shown in Figure~3. Note that for galaxies with
high $S/N$, like the Treu et al.\ sample, the constraints in both age
and metallicity are very tight and thus the so-called age-metallicity
degeneracy is dramatically reduced and does not affect the age
determination at a significant level.

Before we can use the estimated ages to constrain $H_0$ we have to
estimate the effects of the potential systematic uncertainties related to
adopting a single-burst stellar population model and a single-metallicity
model. As far as the first uncertainty is concerned, we know that even
massive field E/S0 galaxies can experience some level of star formation
activity at $z>0.2$ \citep{MAE01,TSCMB02,WHWL02}. The presence of a young
and luminous stellar population superimposed on a dominant old stellar
population can bias the single-burst equivalent ages of the spectrum
to be much younger than the mass-weighted average age. Although this
is a serious concern for the interpretation of individual cases, our
method relies on the old envelope of the age distribution at any given
redshift. Therefore, the effects of this bias on the determination
of $H_0$ and $w$ are strongly reduced, as we discuss further in
\S~3.2. Concerning single metallicity, the uncertainty connected to
metallicity gradients is negligible with respect to other sources of
uncertainty. In fact, our spectra typically cover $6$ kpc (for the LRG)
to $10$ kpc (for the higher redshift samples).  Given that typical
metallicity gradients in E/S0s follow $d {\rm log} [{\rm Fe}/{\rm H}]/
d {\rm log} r = -0.2$ \citep{Davies+93}, metallicity only changes by 0.2
dex within this region. This is fairly small and the single-metallicity
approximation affects our results well within the 1$\sigma$ value of
the error in the recovered metallicity.

\subsection{An age--redshift relation and $w(z)$}

Figure~4 shows the derived single-burst equivalent ages of the galaxies
as a function of redshift.  The circles correspond to galaxies in the SDSS
LRG sample, triangles to the Treu et al.\ sample, diamonds are galaxies in
MS1054$-$0321, and crosses are 53W091 and 53W069.  For clarity, the SDSS
LRG points have not been plotted for ages $<7$ Gyr.  Typical errors on
the ages of LRG galaxies are 10\% and are not plotted.  An age--redshift
relation (i.e., an ``edge'' or ``envelope'' of the galaxy distribution
in the age--redshift plane) is apparent from $z=0$ to $z=1.5$.

Although an envelope can be identified, it is important to notice that
there is a large spread of ages at any given redshift. In particular,
the measured age for many galaxies makes them significantly younger
than the age--redshift envelope.  This is in part due to a real spread
in the star formation histories of early-type galaxies, especially in
the field, but mostly a result of using single-burst equivalent ages.
Single-burst equivalent ages effectively measure a luminosity-weighted
age.  Therefore, even a relatively minor episode of star formation is
sufficient to significantly underestimate the age of the bulk of (old)
stars. Indeed, many of the galaxies in the Treu et al.\ sample show
[\ion{O}{2}]~$\lambda$3727 emission, consistent with minor ongoing star
formation activity \citep{TSCMB02}. Is this contamination enough to
bias our results, or is the use of the relative ages of the old-envelope
sufficient to reduce this bias to useful levels?

To answer this question we consider a scenario where the bulk of stars in
E/S0 galaxies formed at high redshift, while a relatively small fraction
(up to 10\% in mass) of stars is formed at later times ($z<1.5$). This
scenario is approximately consistent with the observed evolution of
the number density \citep{Im+02}, colors \citep{JFDTPN99, MAE01}, and
spectral properties of early-type galaxies \citep{Trager+00, TSCMB02}.
Within this scenario, we use Monte Carlo simulations to estimate the
effects of this bias on the age--redshift relation and ultimately on $H_0$
and $w(z)$ determinations.  We generate galaxy spectra assuming that
after the initial burst of star formation at $z>3$ an extra 10\% of the
stellar mass is formed at a time $t_1$ between redshift $0 < z < 1.5$ with
uniform probability for $t_1$. For simplicity, we use discrete values for
the ages with spacing 0.1 Gyr and assume a flat $\Omega_m=0.27$, $w=-1$
and $h=0.71$ cosmology to convert time to redshift. The metallicity for
a given galaxy is chosen to be uniformly distributed in the range $0.1 <
Z/Z_{\odot} < 4$.  Finally, we add random noise to the spectra so that
$S/N=15$ and determine the age with the same procedure used for the data.

The results of our Monte Carlo simulations are shown in Figure~5. The
top panel shows the simulated age--redshift distribution if only six
early-types are found per redshift interval\footnote{We have drawn
galaxies ages randomly from a grid that spans the age range 0.1 to
14~Gyr, with a spacing 0.1~Gyr.  For our standard, baseline cosmology,
this age spacing translates to the redshift interval seen in Figure~4.}.
The edge is blurred and smeared due to recent star formation contamination
and because of the error in the age determination; nevertheless, it is
clearly recognizable. The galaxies observed within a few 100 Myr of
bursts have very small single-burst equivalent ages, as observed for some
galaxies in the LRG and Treu et al.\ sample, but they do not interfere
with the location of the old envelope.

The lower panel shows the distribution for the case where 60 early-types
are observed per redshift bin (this is less than the number of galaxies
in the SDSS LRG sample). The solid line shows the theoretical position
of the edge, which is remarkably close to observed edge.  Note also that
increasing the number of galaxies has improved the agreement between
the observed and theoretical edge. Thus we conclude that small, recent
bursts of star formation do not introduce a significant bias on the edge
position; our primary uncertainty in determining $w(z)$ to high redshift
is the sparse size of the sample of distant, early-types considered here.
We can now attempt to recover $dz/dt$ from this simulation by looking at
the shift in the edge position as a function of $z$ \citep[see \S~3.4
and][]{JL02} and thus deduce $w(z)$.  We recover $w(z)=-1 \pm 0.2$ for
the whole redshift range. This shows that, with a sample similar to the
LRG sample extended to higher redshift, we should be able to measure
$w(z)$ within a few 10\%. This is similar to what was already concluded
in \citet{JL02}.

Having established that the age--redshift relation for the old envelope is
not significantly affected by the bias caused by later episodes of star
formation, we now consider the observational limits that we can impose
on $w(z)$.  As a consistency check we note that the age at $z=0$ obtained
here is in good agreement with GC ages (see \S 2) and that the Treu et al.\
galaxy ages agree with those from of SDSS LRG sample where they overlap.
The solid line in Figure~4 corresponds to the age--redshift relation
for our flat, reference $\Lambda$CDM model: $\Omega_m=0.27$, $h=71$,
and $w=-1$.  This is consistent with the observed age--redshift envelope,
and it seems to indicate that we live in a universe with a classical
vacuum energy density and that, on average, stars
in the oldest galaxies formed about 0.7 Gyr after the Big Bang (e.g.,
at redshifts $z \sim 7-10$). To make the observed age--redshift relation
consistent with a model with widely different $w(z)$ behavior, we would
have to infer that early-type galaxies at different redshifts have very
different formation epochs for their stellar populations.  For example,
the dotted line in Figure~4 indicates the age-redshift relation for an
evolving population in a universe where $w(z)=-2$ for $z > 1$ and grows
linearly from $w=-2$ at $z=1$ to $w=0$ at $z=0$.  For this model to work,
we would have to conclude that the oldest galaxies at $z\sim 1.5$ formed
about 0.7 Gyr after the Big Bang, but that the oldest galaxies at $z\sim
0$ formed $\sim 3$ Gyr after the Big Bang.  Since the oldest galaxies in
the LRG sample are as old as the oldest GCs, we consider this to be an
unlikely explanation.  Furthermore, not only would this scenario require
that {\emph all} local galaxies either formed their stars more recently
than galaxies at high redshift or had a recent episodes of recent star
formation, but would also require a remarkable fine-tuning for them all
to obtain exactly the same age.  We therefore conclude that this extreme
model for $w(z)$ is unlikely given the data.

Unfortunately, the small number of galaxies in our samples at $z>0.2$
does not allow us to compute $dz/dt$ with enough accuracy to constrain
$w(z)$, and therefore this measurement will have to await better data.
However, the $z < 0.2$ region is well populated by LRG galaxies (Figure~4)
and we can determine $dt/dz$ at $z\sim 0$. This is illustrated in the
bottom panel of Figure~4 by the clear shift in the upper envelope of the
age histogram between the two redshift ranges $0 < z < 0.04$ and $0.08 <
z < 0.12$. In the next section we will investigate this phenomenon in
more detail and use it to determine $H_0$.

\subsection{The value of $H_0$}

By concentrating on the age--redshift relation for SDSS LRG galaxies at
$z \la 0.2$, we now determine the Hubble constant $H_0$.  This procedure
relies upon determining the ``edge'' of the galaxy age distribution in
different redshift intervals.

If the ages of the galaxies were known with infinite accuracy, for
each galaxy$_i$ at redshift $z_i$, one could associate an age $a_i$.
The probability for the age of the oldest stellar population in
galaxy$_i$, $P(t|a_i)$, would be given by a step function which jumps
from 0 to 1 at $a_i$.  An age--redshift relation edge could then be
obtained by dividing the galaxy sample into suitably-large redshift bins
and multiplying the $P(t|a_i)$ for all the galaxies in each bin.

In practice, the age of each galaxy is measured with some error
$\delta a_i$.  We thus assume $P(t|a_i)=1$ if $t>a_i+\delta_{a_i}$ and
$\ln P(a_i)=-x^2$ where
$x=(a_i+\delta_{a_i}-t)/(\sqrt{2}\delta_{a_i})$ otherwise.  We have
divided the $z < 0.2$ portion of the LRG sample into 51 redshift bins.
For each bin, we obtain $P(z,t)$ by multiplying the $P(t|a_i)$ of the
galaxies in that bin.  We define the ``edge'' of the distribution
$t(z)$ to be where $\ln P(z,t)$ drops by $0.5$ from its maximum
(i.e. $\Delta \ln P=0.5$. We associate an error to this determination
$\delta_{t(z)}$ given by $t_{2}(z)-t(z)$ where $t_{2}(z)$ corresponds
to where $\ln P(z,t)$ drops by $2$ from its maximum (i.e. $\Delta \ln
P=2$, this approximately corresponds to the 68\% confidence level).
This procedure makes the determination of the ``edge'' less sensitive
to the outliers (see also \citet{RBBG97}). Figure~6 presents the
resultant $t(z)$ relation.

For $z<z_{\rm max}$, we fit $t(z)$ with a straight line whose slope
$dt/dz$ is related to the Hubble constant at an effective redshift
by $H(z_{\rm eff})= - \frac{1}{1+z_{\rm eff}} (\frac{dt}{dz})^{-1}$.
This fit is performed by standard $\chi^2$ minimization.  We also compute
$P_{\ge \chi^2}$, the probability of obtaining equal or greater value of
the reduced $\chi^2$ if the $t(z)$ points were truly lying on a straight
line.  Values of $P_{\ge \chi^2}<0.1$ means that a straight line is not
a good fit to the points.

For $w = -1$, we obtain $H_0=H(z_{\rm eff})\left[\Omega_m(1+z_{\rm
eff})^3+\Omega_{\Lambda}\right]^{-\frac{1}{2}}$ where $\Omega_m=0.27$,
$\Omega_\Lambda=0.7$. Figure~7 shows how the $H_0$ measurement
depends on $z_{\rm max}$. For $z_{\rm max} > 0.17$ not only the value
for $H_0$ drifts but also a straight line is not a good fit to the
points ($P_{\ge \chi^2}$ suddenly drops below $0.1$). For our $H_0$
determination, $z_{\rm max}=0.17$, $P_{\ge \chi^2}=0.32$, $z_{\rm
eff}=0.09$, and the correction from $H(z_{\rm eff})$ to $H_0$ is a
$4\%$ effect. We obtain $H_0=69\pm 12$ km s$^{-1}$ Mpc$^{-1}$.

Our $H_0$ determination is in good agreement with the Hubble Key Project
measurement \citep[$h = 0.72 \pm 0.03 \pm 0.0 7$;][]{Freedman+01},
with the valued derived from the joint likelihood analysis of {\it
WMAP} + 2dFGRS + Lyman-$\alpha$ forest power spectrum \citep[$h =
0.71^{+0.04}_{-0.08}$;][]{SpergelWMAP03}, with gravitational lens
time delay determinations \citep[$h = 0.59^{+0.12}_{-0.07} \pm
0.03$;][]{TreuKoopmans02}, and Sunyaev-Zeldovich measurements \citep[$h =
0.60^{+0.04+0.13}_{0.04-0.18}$;][]{Reese+02}. We emphasize that our method
is fully independent from any of the above measurements; it is reassuring
that so many different methods yield such a consistent value for $H_0$.

\section{Conclusions}

We have obtained tight constraints on the effective value of $w$ using
solely the determination of the ages of the oldest Galactic globular
clusters and the observed position of the first acoustic peak of the
CMB power spectrum.  Our constraint on $w$ is in good agreement with
that of \citet{SpergelWMAP03}, but is subject to different possible
systematics and is very weakly model dependent.  Our constraint is
slightly less stringent than that obtained by \citet{SpergelWMAP03}
from a joint likelihood analysis of {\it WMAP} with six external data
sets ({\it WMAP} + CBI + ACBAR + 2dFGRS + Lyman$\alpha$ forest power
spectrum + Type~Ia supernovae + {\it HST} Key Project $H_0$
determination), but is tighter than the CMB-only ({\it
WMAP}+CBI+ACBAR) determination and comparable to the {\it WMAP} +
ACBAR + CBI + {\it HST} constraint \citep{SpergelWMAP03}.  We
speculate that the joint constraint from {\it WMAP} and all the
external data sets considered in \citet{SpergelWMAP03} could be
further improved by adding stellar populations age determination as an
additional external data set. We have restricted our analysis to
values of $w > -1$ since theoretical models for $w < -1$ seem
difficult to construct \citep{Carroll+03}. The use of external data
sets (see e.g. \citet{KLSW02} and references therein for other
approaches to measure $w$) could help to remove this restriction and
constrain $w$ for values $< -1$.

We have dated a large sample of galaxies and shown that a clear
age--redshift envelope exists.  By computing the relative ages of SDSS
red galaxies at $z < 0.2$ we obtain an estimate of $H_0$ independent of
and in good agreement with determinations from the Hubble Key Project,
gravitational lensing time delays, Sunyaev-Zeldovich, and {\it WMAP}.
We also show that when surveys with a substantial number of high $S/N$
spectra of old galaxies at $z>>0$ become available, a determination
of the possibly time-dependent equation of state parameter $w(z)$ will
be obtainable.

\acknowledgements We thank David Spergel for discussions that stimulated
this investigation. We gratefully thank Fiona Harrison and Megan Eckart
for providing the time during which the spectra of MS1054$-$0321 were
obtained. We also thank Peter Eisenhardt and Charles Lawrence for useful
discussions and Hiranya Peiris for insightful discussion on the position
of the peak and $w$. LV thanks David Schelegel for help with the SDSS
early data release catalogue. The work of RJ is supported in part by
NSF grant AST-0206031. LV is supported by NASA through {\it Chandra}
Fellowship PF2-30022 issued by the {\it Chandra X-Ray Observatory} Center,
which is operated by the Smithsonian Astrophysical Observatory for and on
behalf of NASA under contract NAS8-39073.  The work of DS was carried out
at Jet Propulsion Laboratory, California Institute of Technology, under
a contract with NASA.  The work of TT is partially supported by STScI
grant AR-09222.  Finally, the authors wish to recognize and acknowledge
the very significant cultural role and reverence that the summit of
Mauna Kea has always had within the indigenous Hawaiian community.
We are most fortunate to have the opportunity to conduct observations
from this mountain.

%\bibliographystyle{../STY/apj.bst}
%\bibliography{../STY/raul}

% FIGURES

\begin{figure}
  \plotone{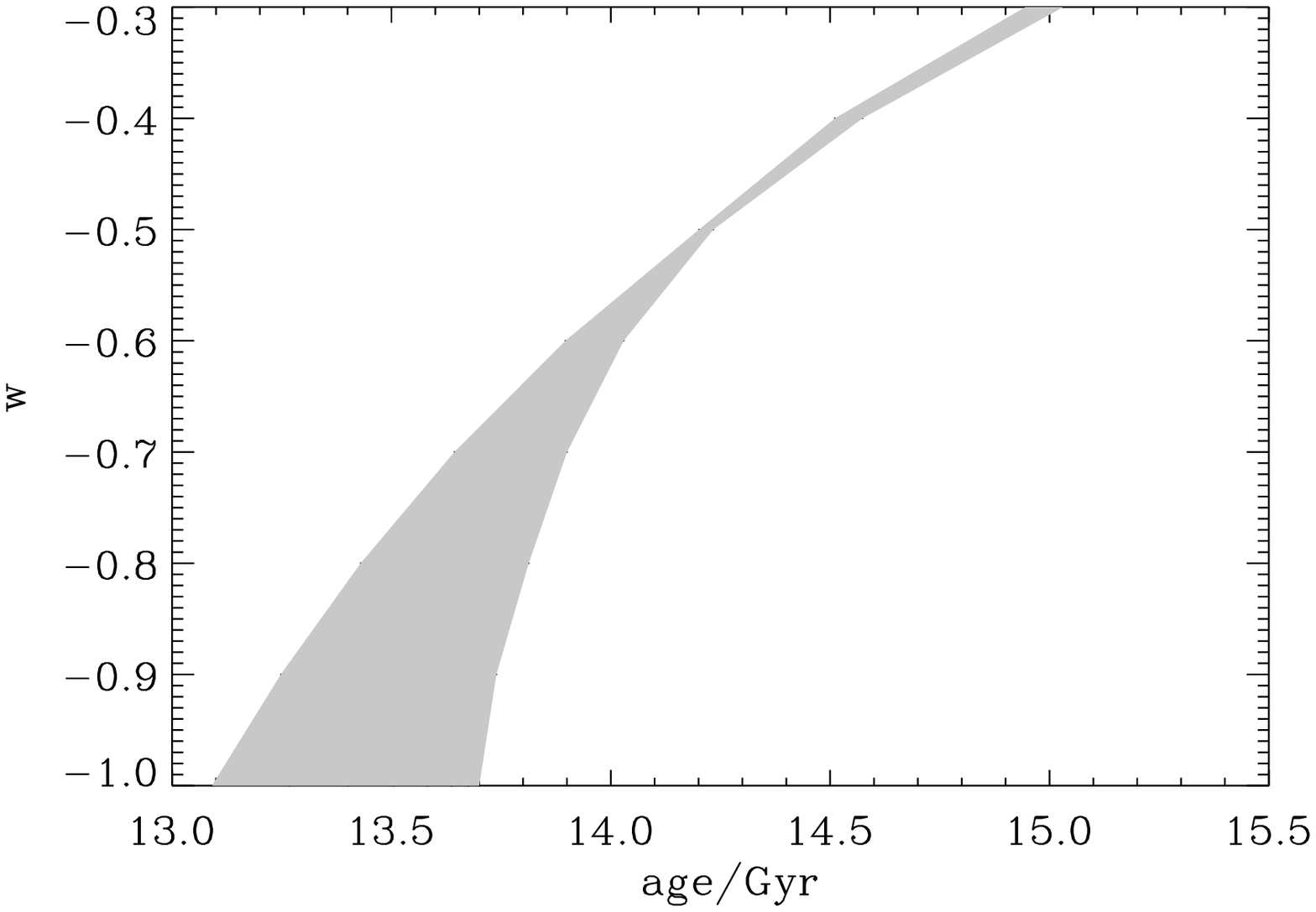}\figcaption{The region of parameter space
    allowed if we fix $\theta_A$ (which in turns fixes the position of
    the first CMB acoustic peak) and $\omega_b \equiv \Omega_b h^2 = 0.02$,
    weakly constraining $0.05<\Omega_m<0.45$ and $0.5<h<0.9$.}
\label{fig:wageth}
\end{figure}

\begin{figure}
  \plotone{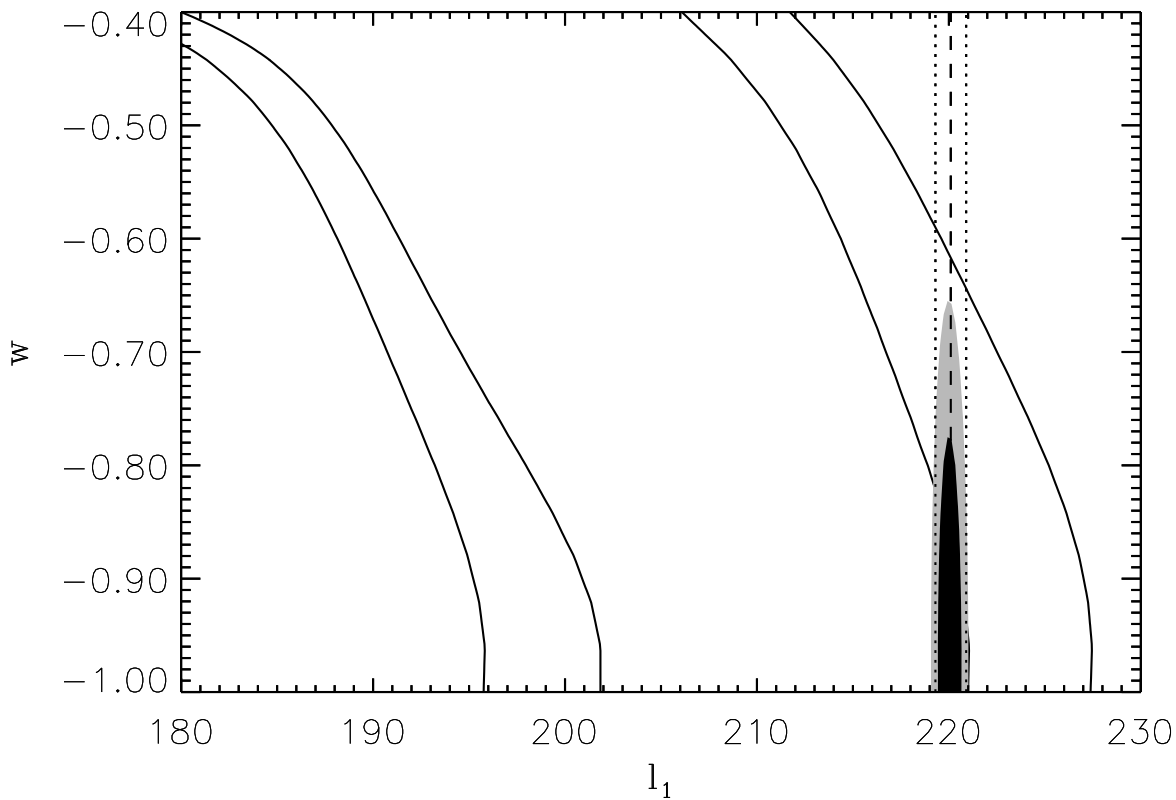} \figcaption{Confidence regions (68 and 90\%)
    for the value of $w$ are shown by the filled grey areas
    (marginalized over the position of the first acoustic peak in the
    CMB, $\ell_1$).  The solid lines denote contours which show the
    constraints in $w$ using the derived age of the oldest Galactic globular
    clusters only. The dash and dotted lines are the the position of
    the first acoustic peak and confidence region as measured by {\it
    WMAP}. Note that the position of the first peak greatly constrains
    the value of $w$.}
\label{lwplane}
\end{figure}

\clearpage
\begin{figure}
  \plotone{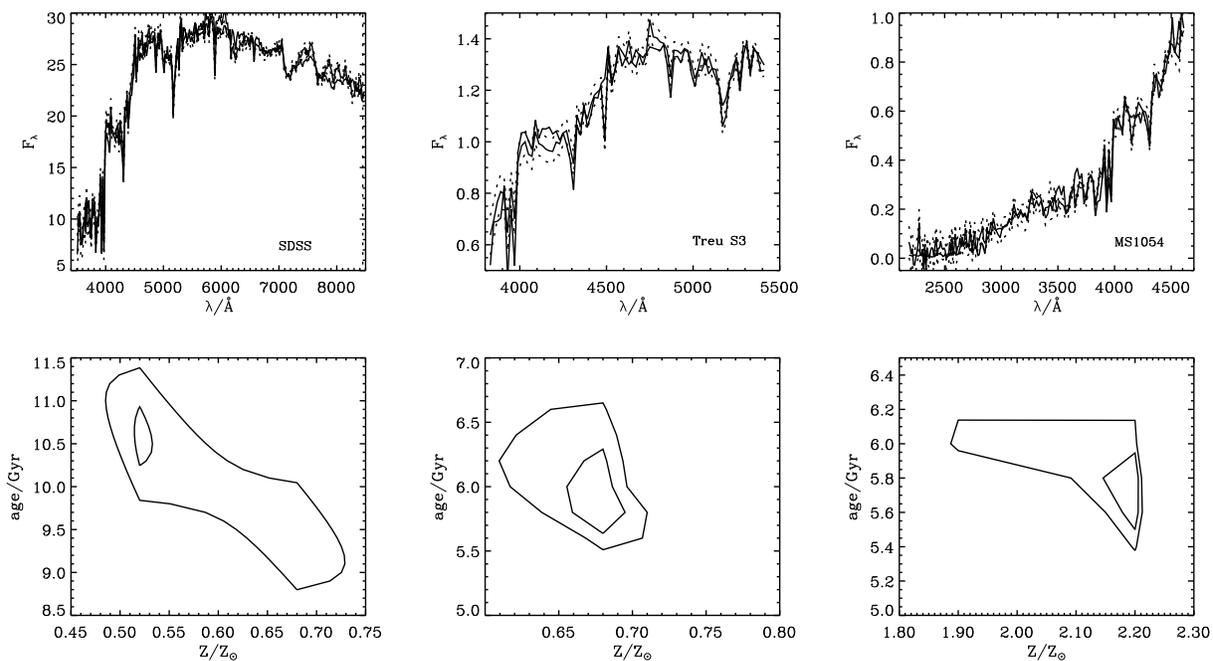} \figcaption{The top 3 panels show examples of
    spectra from the SDSS LRG, Treu et al.\ and MS1054$-$0321 samples. The
    dashed lines correspond to 1$\sigma$ error bars in the flux, the
    thin solid line is the measured spectrum and the thick solid line
    is the best fit to a single stellar population. The 3 bottom
    panels show confidence contours (68 and 95.4 \%) in the recovered
    parameters.}
\label{spectra}
\end{figure}

\clearpage

\begin{figure}
  \plotone{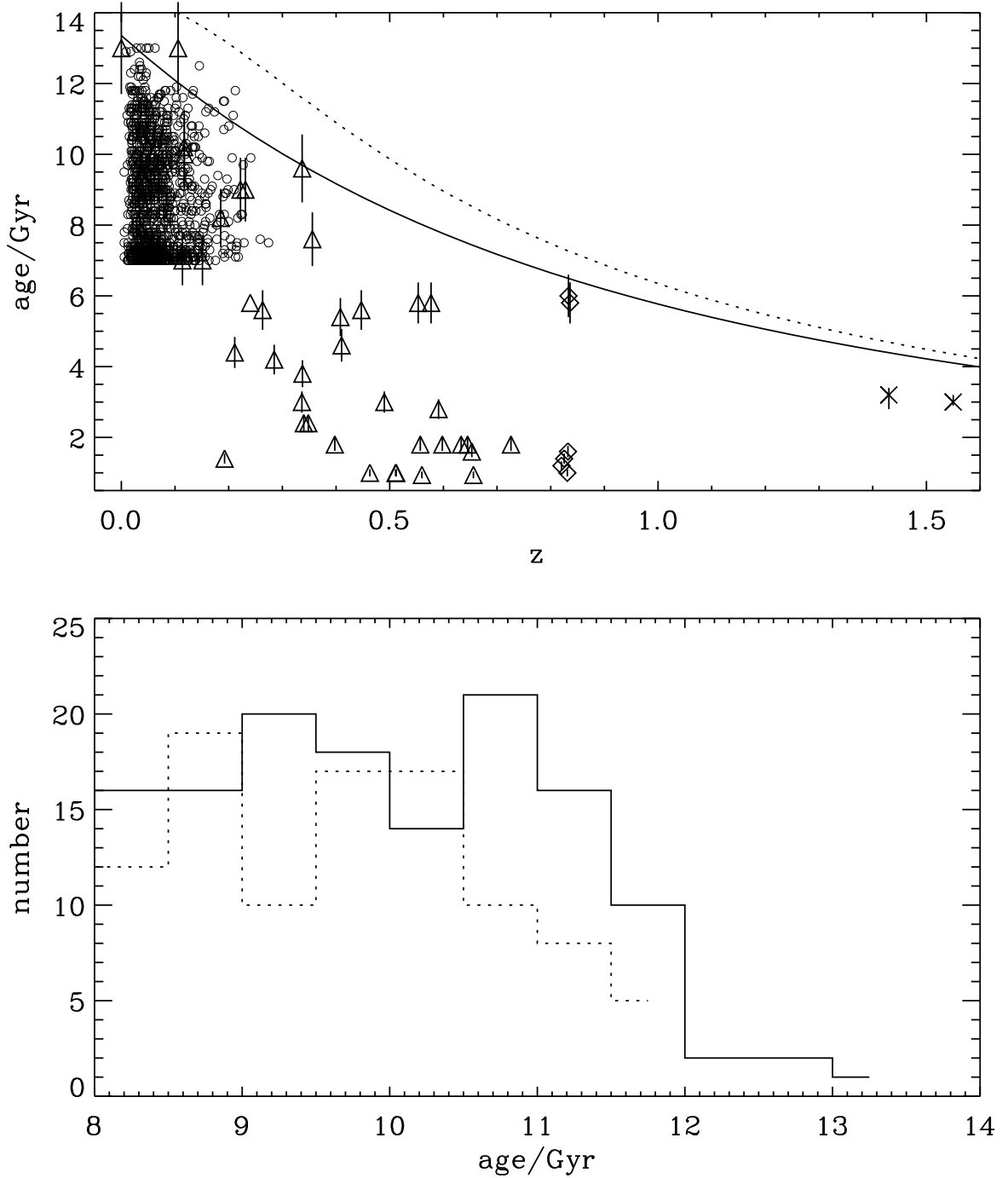} \figcaption{Top panel: Age--redshift envelope
    obtained for the galaxies studied in this paper. Open circles
    correspond to galaxies from the SDSS LRG sample, triangles to the
    Treu et al.\ sample, diamonds to MS1054$-$0321, and crosses are
    53W091 and 53W069.  A clear trend is present: galaxies age as the
    redshift decreases.  The overall shape of this trend is in fair
    agreement with theoretical expectations for a $\Lambda$CDM (solid
    line) with $\Omega_m=0.27$ and $H_0=71$ km s$^{-1}$ Mpc$^{-1}$.
    The dotted line represent an alternative model disfavored by the
    data: $w(z)=-2$ for $z > 1$ and then grows linearly from $w=-2$ at
    $z=1$ to $w=0$ at $z=0$. Bottom panel: histogram along the age
    axis of top panel for the redshift range $0 < z < 0.04$ and $0.08
    < z < 0.12$. The clear shift between the two histograms is a
    measurement of $dz/dt$ and therefore allows us to measure $H_0$.}
\label{agered}
\end{figure}

\clearpage

\begin{figure}
  \plotone{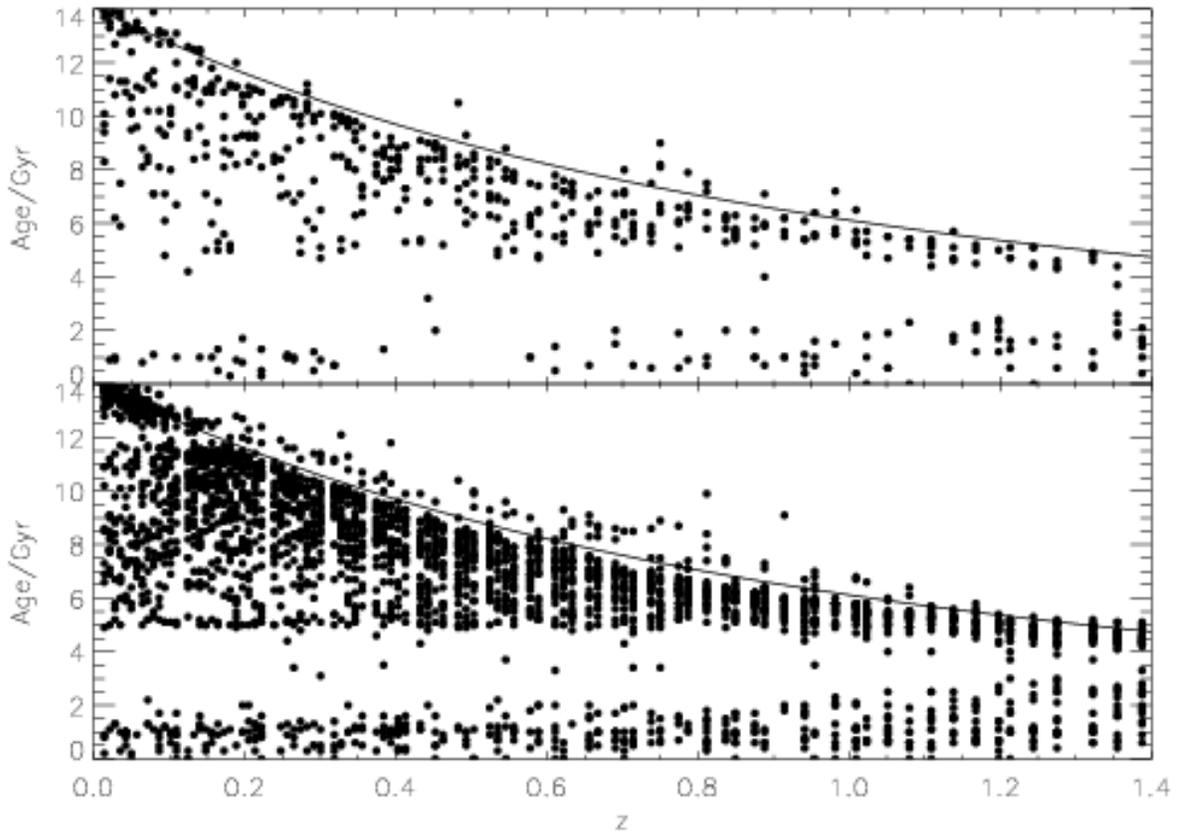} \figcaption{Simulated distribution for the
    age--redshift relation assuming that 10\% of the mass was added to
    the early-type galaxy in the redshift interval $0 < z < 1.5$. We
    simulate two cases: a) there are only six early-type galaxies
    observed per age bin (top panel; similar to the higher-redshift samples),
    and b) there are 60 galaxies per age bin (bottom panel; corresponding 
    to the SDSS sample).  Note that the although the
    edge is less populated due to the fact of recent star formation,
    it is still clearly identifiable in both cases and is not biased.}
\label{mcsimul}
\end{figure}

\clearpage

\begin{figure}
\plotone{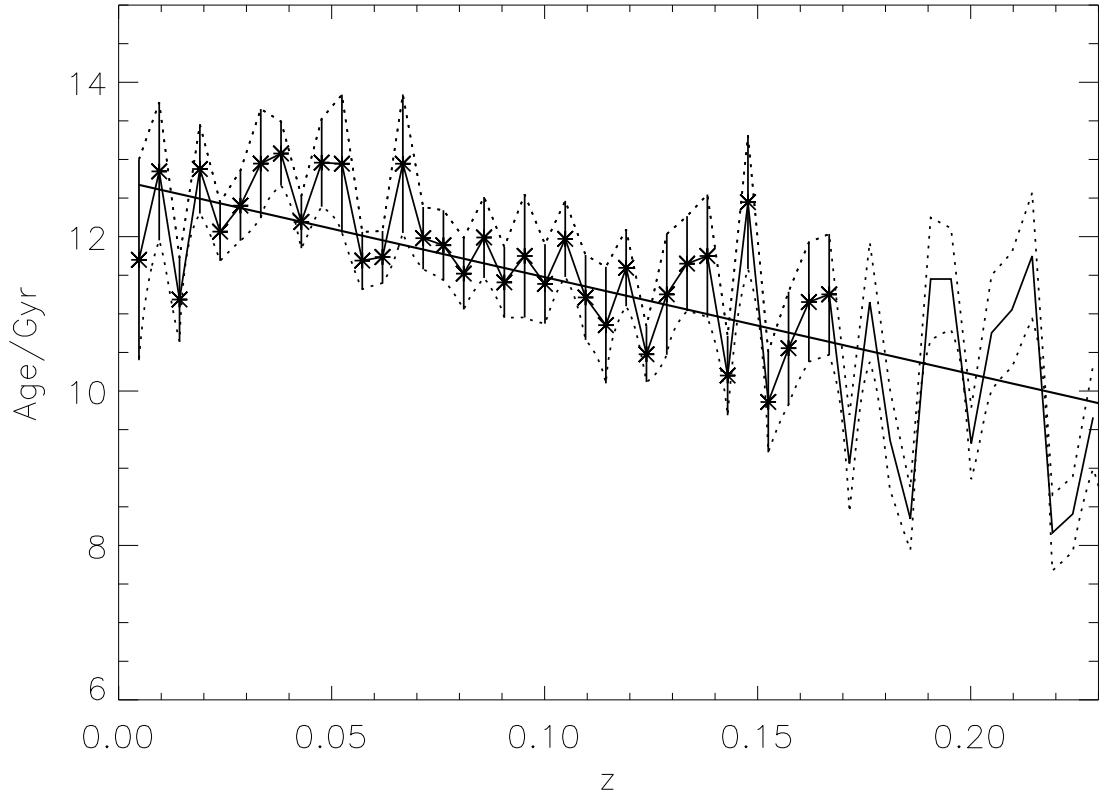} 
\figcaption{The binned, low-redshift age--redshift relation, derived
from the SDSS LRG sample.  The solid line is a best-fit to the edge.}
\label{fig3}
\end{figure}

\clearpage

\begin{figure}
  \plotone{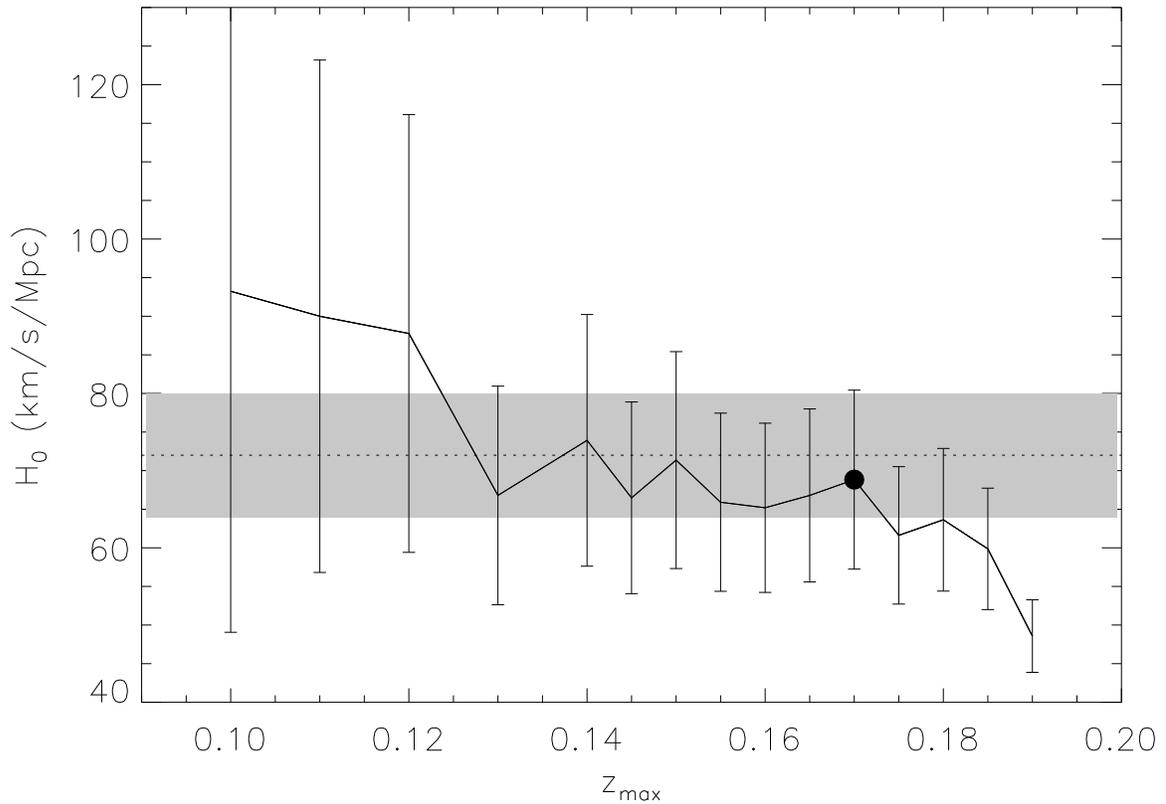} \figcaption{The value for $H_0$ derived from
    Figure~\ref{fig3} as a function of redshift cut-off for computing
    the slope. Note that it is very weakly dependent on the value of
    this cut-off. The solid circle is our adopted value. The shaded
    are corresponds to the 1$\sigma$ confidence region from the Hubble
    Key Project. For $z>0.18$ the probability of the fit being a
    straight line is $P_{\ge \chi^2}<0.1$, while at $z=0.17$ is
    0.32. (see more details in text).}
\label{fig4}
\end{figure}

\clearpage

\end{document}